\documentclass[preprint,amsmath,amssymb,aip,apl]{revtex4}

\usepackage{graphicx}
\usepackage{dcolumn}
\usepackage{bm}

\usepackage{amsmath}
\usepackage{amssymb}
\usepackage{psfrag}
\usepackage{color}
\usepackage[normalem]{ulem}
\usepackage{cancel}

\newcommand{\U}{{\cal U}}

\newcommand{\ws}{w_s}
\newcommand{\Ws}{W_s}

\newcommand{\lec}{\ell_{ec}}

\newcommand{\df}[2]{\frac{\id^4 #1}{\id #2^4}} 

\newcommand{\id}{\mathrm{d}} 

\newcommand{\be}{\begin{equation}}
\newcommand{\ee}{\end{equation}}

\let\baraccent=\= 
\renewcommand{\=}[1]{\stackrel{#1}{=}} 

\begin{document}
\title{The sensitivity of Graphene `Snap-through' to substrate geometry}

\author{Till J.~W.~Wagner}
\affiliation{Department of Applied Mathematics and Theoretical Physics, University of Cambridge,\\ Wilberforce Rd, Cambridge, CB3 0WA, UK }
\author{Dominic Vella}
 \affiliation{OCCAM, Mathematical Institute, University of Oxford, 24-29 St Giles',\\ Oxford, OX1 3LB, UK}

\date{\today}

\begin{abstract}
We study theoretically the deposition of Few Layer Graphene sheets onto a  grooved substrate incorporating adhesion between substrate and sheet. We develop a model to understand the equilibrium of the sheet allowing for partial conformation of sheet to substrate. This model gives new insight into recent observations of `snap-through' from flat to conforming states and emphasizes the crucial role of substrate shape in determining the nature of this transition. Our analytical results are consistent with numerical simulations  using a van der Waals-like interaction . Finally we propose a novel substrate shape that should exhibit a continuous, rather than `snap-through', transition.
\end{abstract}


\maketitle

\begin{figure}[h]
\psfrag{z}[c]{$z$}
\psfrag{x}[c]{$x$}
\psfrag{W}[c]{$z=w(x)$}
\psfrag{l}[c]{$\lambda l$}
\psfrag{s}[c]{{$z=\ws(x)$}}
\psfrag{d}[c]{$\delta$}
\psfrag{h}[c]{$l$}
\psfrag{1}[c]{nonconformal}
\psfrag{2}[c]{partially}
\psfrag{3}[c]{conformal}
\psfrag{4}[c]{conformal}
\setlength\fboxsep{5pt}
\setlength\fboxrule{0.5pt}
\fbox{ \hspace{-3 pt}\includegraphics[width=.99\linewidth]{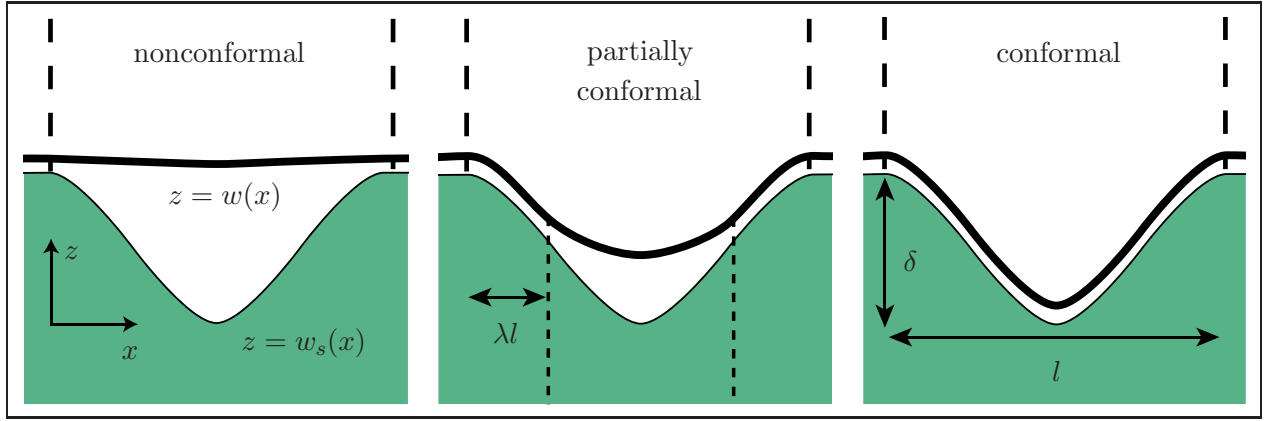}}
\caption{(Color online) Schematic representation of the three possibilities for FLG morphology on a grooved substrate. \textit{Left:} no adhesion and the FLG sheet remains approximately flat, \textit{Centre:} partial adhesion in which the FLG sheet is out of contact with the substrate in the region $ \lambda l< X < (1-\lambda) l$, \textit{Right:} complete adhesion with FLG sheet conforming to substrate morphology.}
\label{fig:setup}
\end{figure}

Recently, considerable research effort has focussed on characterizing the mechanical properties of Few Layer Graphene (FLG) sheets --- from their elasticity and strength\cite{poot08,lee08} to their ultrastrong adhesion to substrates \cite{koenig11,kusminskiy11}. Understanding these properties is important for a number of potential applications since they influence the form taken by sheets, which in turn can influence its electrical characteristics\cite{kim08}. While standard adhesive tests have been applied with some success to measure  adhesion, it has also been proposed that deposition onto a corrugated substrate may provide a simpler assay \cite{zhang11}.  For a given substrate geometry and material properties, we expect that a sheet deposited onto such a substrate will adopt one of  three configurations (illustrated schematically in figure \ref{fig:setup}). For relatively weak adhesion, we expect the sheet to sit above the substrate with very little deflection (the nonconformal\cite{aitken2010,gao2011} scenario in figure \ref{fig:setup}). For very strong adhesion, we expect the sheet to be significantly deflected and to adopt essentially the form of the substrate (the conformal\cite{aitken2010,gao2011} scenario shown in figure \ref{fig:setup}). Previously, it has been assumed that the transition from nonconformal to conformal morphologies is sudden, leading to this transition being referred to as `snap-through'. Indeed, such a snap-through transition has recently been observed in FLG sheets \cite{scharfenberg12}. However, in principle a third, intermediate, morphology exists, which we term `partially conformal' (see figure \ref{fig:setup}), with the sheet conforming to the substrate over a finite portion of its length but not everywhere. In this Letter, we study the transition from nonconformal to conformal morphologies theoretically. The question of principal interest is whether this transition is sudden (i.e.~occurs at a critical adhesive strength) or, rather, whether there is a range of adhesive strengths for which a partially conformal morphology may be observed.

We model the FLG sheet as an elastic beam with bending stiffness $B$ and thickness $h$ whose position is given by $z=w(x)$. To allow for analytical progress we consider a substrate with a single two-dimensional groove of width, $l$, and depth, $\delta$; the detailed shape of the groove is given by $z=\ws(x)$. To determine whether the sheet is conformal, we must determine the shape of the sheet $w(x)$ and compare this with $\ws(x)$. In regions where the sheet is conformal with the substrate we have, by definition, that $w(x)=\ws(x)$. In regions where the sheet is not conformal with the substrate its shape satisfies the beam equation \cite{landau}
\begin{equation}
B\df{w}{x}=0
\label{eqn:beam}
\end{equation} where, for simplicity, we neglect the possibility of a tension within the membrane. (The neglect of the tension within the FLG sheet amounts to neglecting a frictional interaction with the substrate; the inclusion of such a friction would complicate the analysis and require the \emph{ad hoc} assumption that the sheet be on the point of sliding everywhere.)  We assume that the groove is symmetric about $x=l/2$ and denote the position of the contact points between sheet and substrate by $x=\lambda l$ and $x=(1-\lambda)l$. The shape of the FLG sheet is thus
\begin{equation}
w(x)=\begin{cases}
a_0+a_2 (l/2-x)^2,\quad|l/2-x|<\lambda l\\
\ws(x),\quad\quad\quad \quad\quad|l/2-x|>\lambda l
\end{cases}\label{eqn:genshape}
\end{equation} where the constants $a_0$, $a_2$ and the value of $\lambda$ are determined by boundary conditions that we will discuss shortly.

We envisage that the deflection of the free portion of the sheet is caused by the adhesive interaction energy per unit area, $\gamma$, between the substrate and sheet. In particular, we expect that the value of $\lambda$ (which determines the contact points) will be determined to minimize the energy of the system. This energy, $U$, comprises of the bending energy of the sheet (caused by its curvature, which drives the sheet to remain nonconformal) and the energy released by the sheet coming into contact with the substrate over a portion of its length, which drives the sheet towards being conformal. We have that
\be
U = \int_0^{\lambda l} \left[w''(x)\right]^2 \id x + \int_{\lambda l}^{l/2} \left[\ws''(x)\right]^2 \id x - 2\lambda l\gamma, \label{en1}
\ee 
where we take the nonconformal state as the ground state of energy. Using the calculus of variations, it can be shown \cite{landau, majidi07} that the solution $w(x)$ that minimizes the energy \eqref{en1} is given by the solution of the beam equation \eqref{eqn:beam} subject to the boundary condition that\footnote{See Supplementary Material at ... for details of the theoretical analysis.}
\be
w'' (\lambda l)-\ws''(\lambda l) = \sqrt{2 \gamma/B} =\sqrt{2}/\lec,
\label{eqn:landaupeel}
\ee
where  $\lec=(B/\gamma)^{1/2}$ is the \emph{elasto-capillary length} \cite{bico04}. The continuity of sheet displacement and slope at the contact point give the constants $a_0$ and $a_2$ in \eqref{eqn:genshape} as
\begin{eqnarray*}
a_0&=&\ws(\lambda l)-\beta l^2(1/2-\lambda)^2, \\
a_2&=&-\ws'(\lambda l)/l(1-2\lambda),
\end{eqnarray*} 
which may be substituted into \eqref{eqn:landaupeel} to give a single equation for $\lambda$ for given values of the substrate geometry.

In what follows, it will be useful to rescale vertical dimensions by the depth of the substrate groove, $\delta$ and horizontal ones by its width, $l$, i.e.
$
W = w/\delta, \ X = x/l, 
$
etc.. This non-dimensionalization introduces (via \eqref{eqn:landaupeel}) the dimensionless strength of adhesion
\be
\Gamma =(l^4/\delta^2)  \gamma/B = (R/\lec)^2,
\ee where $R=l^2/\delta$ is the typical radius of curvature of the substrate. In physical terms the parameter $\Gamma$ tells us whether the adhesive energy is strong enough to overcome the bending energy penalty resisting the sheet conforming to the substrate. It still remains to be seen, however, whether, for a given substrate shape $\Ws(X)$ the  transition between nonconformal ($\lambda$ = 0, small $\Gamma$) and conformal ($\lambda = 1/2$, large $\Gamma$) is smooth or, rather, a discontinuous `snap-through' transition. We couch our study in terms of varying the dimensionless adhesion strength $\Gamma$, which may be varied by fixing the bending stiffness $B$ and varying $\gamma$ or by holding $\gamma$ constant and varying $B$\cite{evans09}. In the case of FLG sheets the latter approach has been achieved experimentally by varying the number of molecular layers\cite{scharfenberg12}.

\begin{figure}[h]
\psfrag{R}[c]{$\mathcal{W}$}
\psfrag{x}[c]{$X$}
\psfrag{l}[c]{$\lambda$}
\psfrag{y}[c]{$W_s$}
 \includegraphics[width=0.95\linewidth]{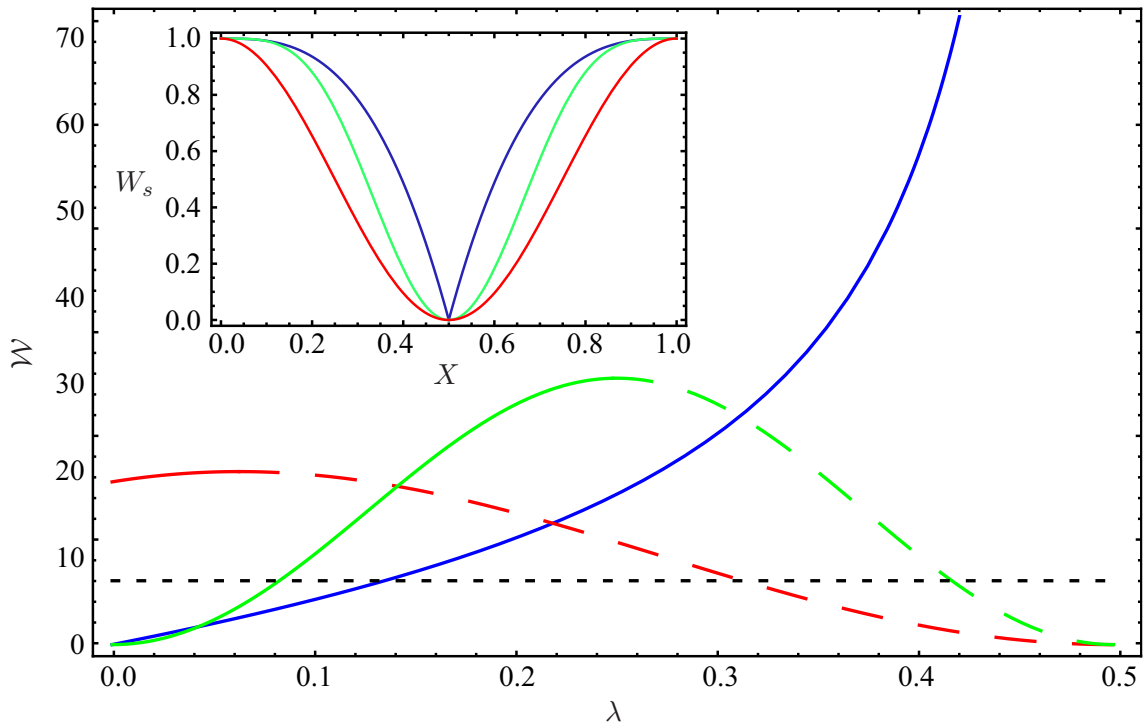}
\caption{Main: The function ${\cal W}(\lambda)$ given by \eqref{eqn:gam_lambda} for the three different substrate shapes considered here: $\Ws^{(1)}$ (red), $\Ws^{(2)}$ (green) and $\Ws^{(3)}$ (blue). The intersection of these curves with the horizontal black line, $(2\Gamma)^{1/2}$, illustrates the values of $\lambda$ satisfying \eqref{eqn:gam_lambda}; these points correspond to maxima (dashed curves) and minima (solid curves) of the  energy. Inset: the three substrate shapes (colors as in main figure). }
\label{three_subs}
\end{figure}

We note that the dimensionless version of the boundary condition eq.~\eqref{eqn:landaupeel}, may be written explicitly in terms of the substrate geometry as
\be
(2\Gamma)^{1/2} = -\Ws''(\lambda)-2\Ws'(\lambda)/(1-2\lambda) \equiv \mathcal{W(\lambda)} .
\label{eqn:gam_lambda}
\ee
For a given substrate shape $\Ws(X)$ and dimensionless adhesion strength $\Gamma$ we therefore have a single equation for $\lambda$ (the position of the contact points). Here we shall consider three substrate morphologies to illustrate some of the different behaviors that can be observed:
\be
\Ws=\begin{cases}\Ws^{(1)}=\tfrac{1}{2}\left(1+\cos2\pi X\right),\\
\Ws^{(2)}=1-\sin^4\pi X,\\
\Ws^{(3)}=1- \left(1-|\frac{1}{2}-X| \right)^3,
\end{cases}
\label{eqn:3subs}
\ee 
for $0\leq X\leq 1$ (outside this interval $\Ws=1$ in each case). For each of these substrate shapes it is a simple matter to plot the behavior of the RHS of \eqref{eqn:gam_lambda} as a function of $\lambda$ (fig.~\ref{three_subs}). We see that for a given value of $\Gamma$ there are typically $0, 1$ or $2$ values of $\lambda$ that satisfy \eqref{eqn:gam_lambda}. The different behaviors shown in figure \ref{three_subs} influence the nature of the transition from nonconformal to conformal morphologies, as we shall see shortly.

The roots of equation \eqref{eqn:gam_lambda} correspond to the extrema of the energy. However, as well as a minimum of energy (corresponding to the contact point we would expect to observe experimentally), there may also be other extrema i.e.~maxima and inflection points. Eqn \eqref{eqn:gam_lambda} does not contain information as to which of its solutions correspond to minima and which are rather maxima or inflection points. In principle, it is possible to obtain this information from the variational approach, using the second variation\cite{gelfand}. However, it is also possible for global energy minima to occur with $\lambda=0$ or $\lambda=1/2$; since these are not extrema they do not appear as solutions of \eqref{eqn:gam_lambda} and must be detected by considering the dimensionless energy $\U=U/(B \delta^2/l^3)$, with $U$ as in \eqref{en1}. We therefore detect the nature of all extrema by considering $\U$.

We shall shortly discuss the different equilibrium states of an FLG sheet for different substrates using the formulation above and considering the value of $\lambda$ for which $\U$ is minimized. However, an alternative approach, which has been adopted in related studies \cite{aitken2010,gao2011,bodetti12}, is to model the molecular forces of adhesion directly by means of a medium range attractive and short range repulsive van der Waals force. For a thin sheet resting on a nonpolar substrate the interaction energy between a sheet molecule and a substrate molecule is given by\cite{israelachvili}
\be
U_{m-m} = -C \left( r^{-6} - Dr^{-12} \right), \label{num_mm}
\ee
with $r$ the distance between the two molecules and $C$ and $D$ material dependent parameters. To obtain the full  interaction energy between the FLG sheet and the substrate, \eqref{num_mm} is integrated over the semi-infinite substrate and the full thickness of the beam. To simplify the resulting expression, we assume that the typical slope of the substrate, $\delta/l\ll1$. For a sheet of thickness  $H = h/\delta$ and with the distance between the mid-plane of the sheet and the surface of the substrate denoted by $Y(X) = W(X)-\Ws(X)$ the beam equation \eqref{eqn:beam} is modified to become in dimensionless form\cite{Note1}
\begin{align}
0= \df{W}{X} + &\alpha \left\{(Y-H/2)^{-3} - (Y+H/2)^{-3} - \right. \nonumber \\
&\left. \beta \left[ (Y-H/2)^{-9} - (Y+H/2)^{-9} \right] \right\} . \label{nums_equ}
&\end{align} The dimensionless constant $\beta$ is related to the equilibrium distance, $Y_0$, between an undeformed sheet and a flat substrate.  We define the distance between the bottom surface of the sheet and the surface of the substrate in this equilibrium as  $Y^* = Y_0 - H/2$. In experiments presented previously\cite{scharfenberg12}, $h\gtrsim 6\mathrm{~nm}$, $y^*\lesssim 3.3 \mathrm{~\AA}$\cite{giovannetti08} and $\delta=120\mathrm{~nm}$ so that $Y_0\ll1$, $Y^*\ll H\ll1$. In the limit $H  \gg Y^*$ we find that $\beta \simeq  {Y^*}^6$. Furthermore the constant $\alpha$ and the adhesive energy $\Gamma$ are related by $\alpha \simeq -(8/3)H^2 \Gamma$ in this limit. (Note that the limit $Y_0\gg1$ leads to a different simplification of the model, which is appropriate for substrates with small scale roughness\cite{aitken2010,gao2011} rather than the larger  grooves considered here.) The ordinary differential equation \eqref{nums_equ} can be solved subject to free-end boundary conditions using, for example, the MATLAB boundary value problem solver \texttt{bvp4c}. The results of these numerical simulations can be compared with those of the analytical model by examining the total energy $\U$. We shall see that the results of this analysis and the semi-analytic approach outlined above are generally in good agreement in the limit $1 \gg H \gg Y^*$.

Having outlined our analytical and numerical approaches, we now consider the snap-through characteristics of the three different substrates given in \eqref{eqn:3subs}. We shall see that each of these substrates illustrates a different type of  transition; indeed,  the transition from nonconformal to conformal may be smooth and not a `snap-through' at all.

\begin{figure}[h]
\psfrag{a}[c]{(a)}
\psfrag{b}[c]{(b)}
\psfrag{c}[c]{(c)}
\psfrag{e}[c]{{\tiny decreasing $Y^*$}}
\psfrag{g}[c]{$\Gamma$}
\psfrag{U}[c]{$\U$}
\psfrag{G}[c]{\tiny{$\Gamma$}}
\psfrag{u}[c]{\tiny{$\U$}}
 \includegraphics[width=0.95\linewidth]{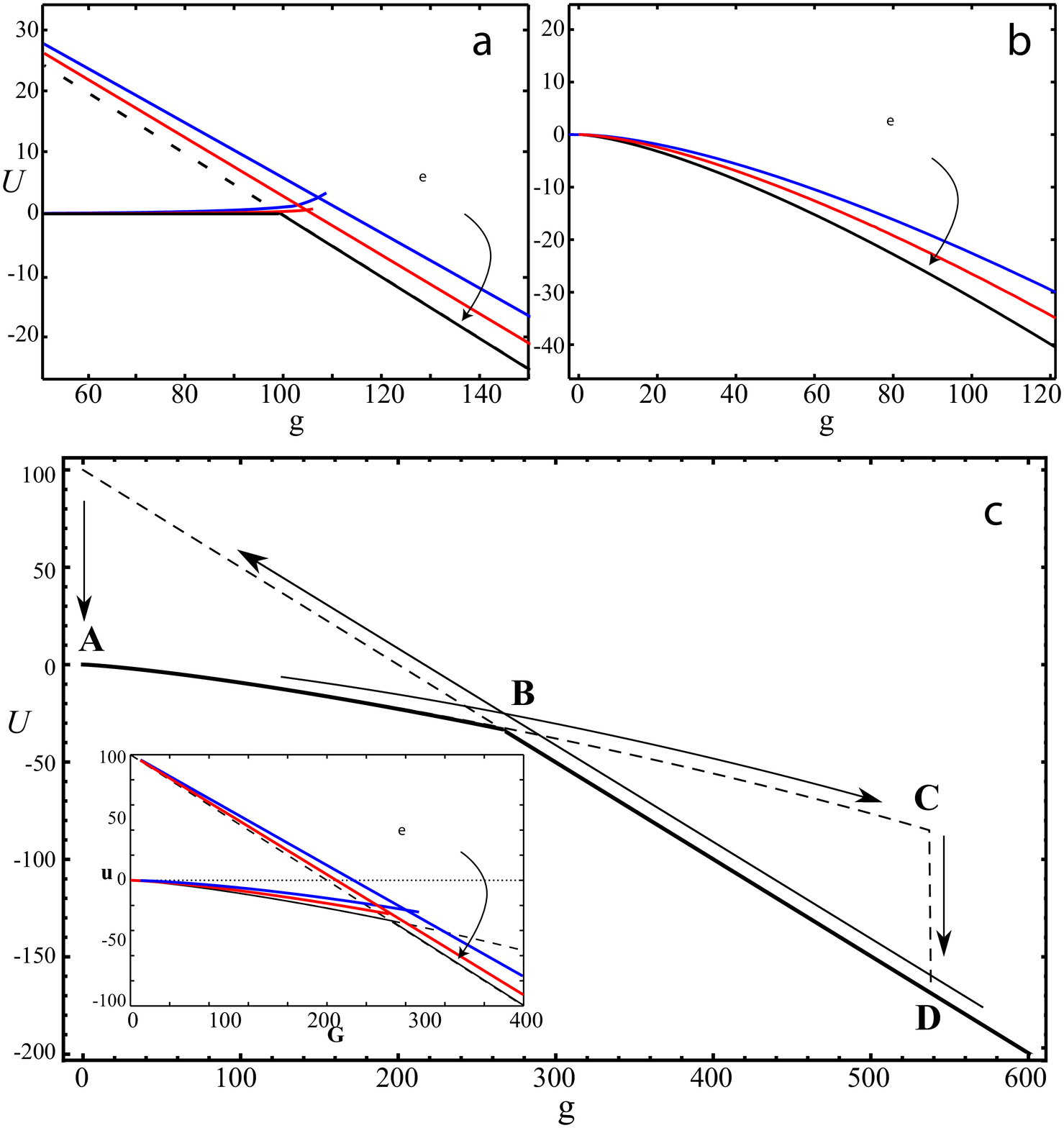}
\caption{(Color online) Energy $\U$ of an FLG sheet above various grooved substrates as a function of adhesion strength $\Gamma$. Results show the predictions based on the analytic model for the global energy minimum (solid black curves) and local energy minimum (dashed curves) along with the numerical results for the van der Waals-like interaction governed by \eqref{nums_equ} with $Y^* = 10^{-3}$ (blue curves) and $Y^* = 10^{-4}$ (red curves), both with $H=0.05$. (a) For $\Ws^{(1)}$ a `snap-through' transition is observed analytically (see discontinuity in slope of black curve at $\Gamma\approx97.4$) and recovered in numerical simulations as $Y^*\to0$. (b) For $\Ws^{(3)}$ a smooth transition from nonconformal to conformal is observed with both approaches. (c) For $\Ws^{(2)}$ partially conformal states are observed before `snap-through' to the conformal state.  Note the onset of a partially conformal morphology (A), earliest snap-through (B) and final snap-through (C) giving rise to hysteresis loop (arrows). Inset: numerical results for $\U$ in this case.}
\label{nums:1} 
\end{figure}

\paragraph*{Fully Discontinuous Snap-Through}

The substrate morphology  $\Ws^{(1)}$ in \eqref{eqn:3subs} is of practical interest since it  closely represents that used experimentally\cite{scharfenberg12}. Examining the corresponding curve for ${\cal W}(\lambda)$ in fig.~\ref{three_subs} we see that only for $2\pi^4 \approx 194.8 \leq \Gamma\lesssim 220.2$ do partially adhered states represent energy minima. From the energy $\U$ for this substrate, we find that a global minimum exists with $\lambda=1/2$ (i.e.~ conformal morphology) for $\Gamma>\pi^4 \approx97.4$ (which is just the bending energy of the conformal state). Since this threshold is significantly below that at which the small window of partially conformal states exists, we expect the transition to the conformal state to be discontinuous, i.e.~a `snap through' occurs. This is confirmed by the numerical results using a van der Waals attraction (see fig.~\ref{nums:1}) and confirms previous assumptions made in the analysis of experimental results\cite{scharfenberg12}.

\paragraph*{Partially Conformal States Before `Snap-Through'}

The substrate morphology  $\Ws^{(2)}$ is qualitatively similar to $\Ws^{(1)}$, albeit with flatter peaks. However, this modification has a significant influence on the behavior of ${\cal W}(\lambda)$ (see fig.~\ref{three_subs}). We see that in this case a partially conformal state exists as the local energy minimum, provided that $\Gamma\leq 529.3$. A calculation of $\U$ shows that this local minimum is the global minimum for $\Gamma\leq257.1$ while for $\Gamma\geq257.1$ the conformal state is the global energy minimum. We thus expect that for $\Gamma\leq257.1$ the sheet will adopt a partially conformal morphology but that for $257.1\leq\Gamma\leq529.3$ the sheet may adopt \emph{either} the partially conformal or the fully conformal state or the fully adhered state. Which of these states is realized in practice depends on the details of the experimental setup, including dynamic considerations. This uncertainty is illustrated by a hysteresis loop in fig.~\ref{nums:1}c, which is also observed in the numerical simulations because of the use of a continuation scheme. Finally, we note that simply determining the adhesion energy required to give $\U=0$, as calculated previously\cite{scharfenberg12}, underestimates the value of $\Gamma$ at snap-through by around $20\%$. This discrepancy arises because for this substrate geometry partially conformal states may have lower energy than the fully conformal state, depending on the value of $\Gamma$. This picture is confirmed by numerical simulations (see inset of fig.~\ref{nums:1}c).

\paragraph*{A Smooth Transition}

The substrate morphology  $\Ws^{(3)}$ is qualitatively different to those of $\Ws^{(1)}$ and $\Ws^{(2)}$ since it has a cusp at the midpoint, $X=1/2$. The behavior of the quantity ${\cal W}(\lambda)$ is also qualitatively different to that observed for other substrates. In particular, we see from fig.~\ref{three_subs} that as $\lambda$ increases towards $\lambda=1/2$, ${\cal W}(\lambda)$ diverges. This divergence means that for all values of $\Gamma$ there is a unique solution of \eqref{eqn:gam_lambda}; in other words, the transition from nonconformal to conformal morphologies progresses smoothly through partially conformal states; no `snap-through' occurs. Physically, this happens because the curvature of the substrate diverges at the cusp and so no finite adhesion energy can overcome the bending energy penalty required to adopt a fully conformal morphology.

In summary, we have presented a new theoretical formulation for the adhesion of FLG sheets onto grooved substrates. This formulation provides a simple explanation for why in the experiments presented to date a discontinuous `snap-through' occurs: the curvature of the peaks of the substrate is so great that the bending energy penalty that has to be paid to conform partially to the substrate is too  large. As a result the sheet can only conform fully and even then only once the adhesion energy is large enough. Our new formulation highlights the crucial role played by the groove geometry in determining the nature of the transition; we have presented a substrate shape for which the sheet conforms partially to the substrate before snapping-through discontinuously and another where the discontinuous nature of the `snap-through' transition disappears entirely to be replaced by a smooth transition from nonconformal to conformal morphologies. Furthermore, we have illustrated that in some cases the adhesion strength at which snap-through occurs may not be determined by setting $\U=0$ since partially conformed states may be energetically favourable. These findings are supported by numerical simulations of a more traditional van der Waals-type model, which converges rapidly to the analytic result in the limit $Y^* \ll H \ll  1$.

Finally, we emphasize that we have proposed a substrate shape, $\Ws^{(3)}$, for which the discontinuous `snap-through' transition is replaced by a smooth family of partially conformal morphologies. This development may allow for the determination of the strength of adhesion from a single experiment, since the position of the contact points encodes information about the strength of adhesion --- it is not necessary to do a whole series of experiments with different thickness FLG sheets as performed previously\cite{scharfenberg12}. While the shape $\Ws^{(3)}$ may seem difficult to fabricate, we note that a qualitatively similar shape is frequently seen at grain boundaries \cite{style05,bouville07}. The continuous adhesion transition of this groove shape may also be of interest for other experiments on thin layer materials; for example, the high curvature seen near the cusp may give rise to plastic deformations making this a simple system within which to study the plasticity of FLG sheets.

This publication is based on work supported in part by Award No.~KUK-C1-013-04, made by King Abdullah University of Science and Technology (KAUST). TJWW~is supported by EPSRC.

\bibliographystyle{apsrev} 
\bibliography{blisters}  

\end{document}